\documentclass[preprint,preprintnumbers,amsmath,amssymb,superscriptaddress,11pt,graphicx]{revtex4}

\usepackage{graphicx}
\usepackage{dcolumn}
\usepackage{bm}
\usepackage{amssymb}

\newcommand{\be}{\begin{equation}}
\newcommand{\ee}{\end{equation}}
\newcommand{\ba}{\begin{eqnarray}}
\newcommand{\ea}{\end{eqnarray}}

\def\mpl{m_{\mathrm{Pl}}}
\def\Mpl{M_{\mathrm{Pl}}}

\begin{document}

\title{Improved calculation of relic gravitational waves}

\author{Wen Zhao}
\affiliation {Department of Physics, Zhejiang University of Technology, Hangzhou, 310014, People's Republic of China}

%\date{\today}

%%%%%%%%%%%%%%%%%%%%%%%%%%%%%%%%%%%%%%%%%%%%%%%%%%%%%%%%%%%%%%%%%%%%%%%%%%%%%%%%%%%
%%%%%%%%%%%%%%%%%%%%%%%%%%%%%%%%%%  ABSTRACT  %%%%%%%%%%%%%%%%%%%%%%%%%%%%%%%%%%%%%%%%%%
%%%%%%%%%%%%%%%%%%%%%%%%%%%%%%%%%%%%%%%%%%%%%%%%%%%%%%%%%%%%%%%%%%%%%%%%%%%%%%%%%%%

\begin{abstract}

In this paper, we improve the calculation of the relic
gravitational waves (RGW) in two aspects: First, we investigate
the transfer function after considering the redshift-suppression
effect, the accelerating expansion effect, the damping effect of
free-streaming relativistic particles, and the damping effect of
cosmic phase transition, and give a simple approximate analytic
expression, which clearly illustrates the dependent relations on
the cosmological parameters. Second, we develop a numerical method
to calculate the primordial power spectrum of RGW at a very wide
frequency range, where the observed constraints on $n_s$ (the
scalar spectral index) and $P_S(k_0)$ (the amplitude of primordial
scalar spectrum) and the Hamilton-Jacobi equation are used. This
method is applied to two kinds of inflationary models, which all
satisfy the current constraints on $n_s$, $\alpha$ (the running of
$n_s$) and $r$ (the tensor-scalar ratio). We plot them in the
$r-\Omega_g$ diagram, where $\Omega_g$ is the strength of RGW, and
study their detection by the CMB experiments and laser
interferometers.

\end{abstract}

%%%%%%%%%%%%%%%%%%%%%%%%%%%%%%%%%%%%%%%%%%%%%%%%%%%%%%%%%%%%%%%%%%%%%%%%%%%%%%%%%%%
%%%%%%%%%%%%%%%%%%%%%%%%%%%%%%%%%%%%%%%%%%%%%%%%%%%%%%%%%%%%%%%%%%%%%%%%%%%%%%%%%%%

%\pacs{ 98.80.-k, 98.80.Es, 04.30.-w, 04.62.+v}

\maketitle

%%%%%%%%%%%%%%%%%%%%%%%%%%%%%%%%%%%%%%%%%%%%%%%%%%%%%%%%%%%%%%%%%%%%%%%%%%%%%%%%%%%
%%%%%%%%%%%%%%%%%%%%%%%%%%%%%%%%%%  SECTION 1   %%%%%%%%%%%%%%%%%%%%%%%%%%%%%%%%%%%%%%%%%%
%%%%%%%%%%%%%%%%%%%%%%%%%%%%%%%%%%%%%%%%%%%%%%%%%%%%%%%%%%%%%%%%%%%%%%%%%%%%%%%%%%%

\section{Introduction}

Recently, a lot of observations on the CMB power spectra
\cite{map1,map3,other} and the large scale structure (LSS)
\cite{lss} have supported inflation as the good phenomenological
model to describe the evolution of the universe at very early
stage, which naturally answers the origin of the primordial
fluctuations with a nearly scale-invariant and gaussian spectrum.
In addition to the density perturbations, inflationary models also
predict a stochastic background of relic gravitational waves
(RGW), which is also called the tensor perturbations. The
amplitude of RGW directly relates to the energy scale of
inflation. Although this background has not yet been observed, and
only some loose constraints have been achieved
\cite{map3,constraint}, but its detection would provide
incontrovertible evidence that inflation actually occurred and
would set strong constraints on the dynamic of inflation
\cite{sasaki}. So it is always regarded as the ``smoking-gun"
evidence for inflation.

There are mainly two kinds of experiments are underway to detect
the RGW at different frequency. One is the CMB experiments, which
can find the RGW by observing the CMB B-polarization power
spectrum \cite{B-Pol}. This method is sensitive to the waves with
very low frequency, $\nu\in(10^{-17},~10^{-15})$Hz. Now, the
first-three-year results of WMAP \cite{map3} have not found the
evidence of the gravitational waves, and only give a constraint
$r<0.28 (95\%$ C.L.), where $r$ is the so-called tensor-scalar
ratio. The next experiment, the Planck satellite \cite{planck},
has higher sensitivity to polarization, which is scheduled for
launch in $2007$, and expects to observe the RGW if $r>0.1$. The
ground-based experiment, Clover (Cl-Observer) is also under
development \cite{clover}, which is expected to observed the RGW
if $r>0.005$. Another kind of important experiments are the laser
interferometers, including the BBO (Big Bang Observer) \cite{BBO}
and DECIGO (DECihertz Interferometer Gravitational wave
Observatory) \cite{DECIGO}, which can detect the gravitational
waves with very high frequency $\nu\sim0.1$Hz. The former can
detect the RGW when $\Omega_g>2\times10^{-17}$ is held, where
$\Omega_g$ is the strength of RGW at $0.1$Hz, and the latter
expects to observe if $\Omega_gh^2>10^{-20}$. It should notice
that the waves with very high frequencies can be observed by the
electromagnetic resonant system\cite{lify}. This is also an
important method to detect the relic gravitational waves.

%So the simple power-law spectrum is a very good approximation for
%the small $r$, but doesn't apply for the case with large $r$.

A lot of works have been done to study the RGW detection by these
experiments \cite{before-work}. In the previous work
\cite{previous}, we have discussed the predicted values of RGW
($r$ and $\Omega_g$) for some kinds of inflationary models, where
we have used a simple power-law function to describe the
primordial power spectrum of RGW, which is a very good
approximation for the waves with very low frequency, but for the
waves with high frequency, this may generate large error. At that
work, we have not considered the damping effect of cosmic phase
transition on the RGW, such as the QCD transition
\cite{trans,qcd,ee}, $e^+e^-$ annihilation and so on. In this
paper, we will discuss this topic more precisely: First, we will
consider the damping effect of a general cosmic phase
transformation, which can been described by a simple damping
factor. And then we will give a simple form of the total transfer
function, which applies to the waves with $\nu\gg10^{-16}$Hz. This
function is dependent on the values of $\Omega_{\Lambda}$,
$\Omega_{m}$, the present energy densities of vacuum and matter,
respectively; the value of $\tau_0$, the age of the universe; the
value of $H_0$, the present Hubble constant; the values of $g_{*}$
and $g_{*s}$, the effective number of relativistic degrees of
freedom when the waves exactly crossed the horizon; and the
fraction $f$ of the background (critical) energy density of the
free-streaming relativistic particles in the universe when the
waves exactly crossed the horizon. So this function includes
abundant cosmic information. Second, we will use a numerical
method to calculate the primordial power spectrum of RGW, where
the Hamilton-Jacobi formula is used. Compared with the simple
power-law form, this numerical result has little change for the
value of $\Omega_g$ when $r$ is smaller, $r<0.02$. But when the
value of $r$ is larger, the numerical result is obviously smaller
than that of the simple power-law approximation.

The organization of this paper is as follows: in section 2, we
will simply review the evolutive equation of the RGW. In section
3, we mainly discuss the damping effects. In section 4, we will
introduce the numerical method by discussing two kinds of
inflationary models. At last, we will give a conclusion and
discussion in section 5.

\section{The Relic Gravitational Waves}

Incorporating the perturbation to the spatially flat
Robertson-Walker (FRW) spacetime, the metric is
 \be\label{1}
 ds^2=a(\tau)^2[d\tau^2-(\delta_{ij}+h_{ij})dx^idx^j]~,
 \ee
where $a$ is the scale factor of the universe, $\tau$ is the
conformal time, which relates to the cosmic time by $ad\tau\equiv
dt$. The perturbation of spacetime $h_{ij}$ is a $3\times3$
symmetric matrix. The gravitational wave field is the tensorial
portion of $h_{ij}$, which is transverse-traceless
$\partial_ih^{ij}=0$, $\delta^{ij}h_{ij}=0$. Since the
gravitational waves are very weak, $|h_{ij}|\ll1$, one needs just
study the linearized evolutive equation:
 \be\label{h_evo_0}
 \partial_{\mu}(\sqrt{-g}\partial^{\mu}h_{ij})=16\pi
 Ga^2(\tau)\Pi_{ij}~,
 \ee
where $\Pi_{ij}$ is the tensor part of the anisotropy stress,
which satisfies $\Pi_{ii}=0$ and $\partial_i\Pi_{ij}=0$, and
couples to $h_{ij}$ like an external source in this equation,
which is always generated by the free-streaming relativistic
particles \cite{damp2,boy}, the cosmic magnetic \cite{magnetic},
and so on. It is convenient to Fourier transform as follows
 \be\label{3}
 h_{ij}(\tau,{\bf x})=\sum_\lambda\sqrt{16\pi G}\int \frac{d~{\bf k}}{(2\pi)^{3/2}}
 \epsilon_{ij}^{(\lambda)}({\bf k})h_{\bf k}^{\lambda}(\tau)e^{i{\bf
 kx}}~,
 \ee
 \be\label{4}
 \Pi_{ij}(\tau,{\bf x})=\sum_\lambda\sqrt{16\pi G}\int \frac{d~{\bf k}}{(2\pi)^{3/2}}
 \epsilon_{ij}^{(\lambda)}({\bf k})\Pi_{\bf k}^{\lambda}(\tau)e^{i{\bf
 kx}}~,
 \ee
where $\lambda=`+`$ or $``\times"$ labels the two polarization
states of the gravitational waves.  The polarization tensors are
symmetry, transverse-traceless $k^i\epsilon_{ij}^{(\lambda)}({\bf
k})=0$, $\delta^{ij}\epsilon_{ij}^{(\lambda)}({\bf k})=0$, and
satisfy the conditions $\epsilon^{(\lambda)ij}({\bf
k})\epsilon_{ij}^{(\lambda')}({\bf k})=2\delta_{\lambda\lambda'}$
and $\epsilon_{ij}^{(\lambda)}({\bf
-k})=\epsilon_{ij}^{(\lambda)}({\bf k})$. Since the RGW we will
consider is isotropy, and each polarization state is same, we can
denote $h_{\bf k}^{(\lambda)}(\tau)$ by $h_k(\tau)$, and $\Pi_{\bf
k}^{(\lambda)}(\tau)$ by $\Pi_k(\tau)$, where $k=|\textbf{k}|$ is
the wavenumber of the gravitational waves, which relates to the
frequency by $\nu\equiv k/2\pi$ (the present scale factor is set
$a_0=1$). So Eq.(\ref{h_evo_0}) can be rewritten as
 \be\label{h-evolution}
 \ddot{h}_{k}+2\frac{\dot{a}}{a}\dot{h}_k+k^2h_k=16\pi
 Ga^2(\tau)\Pi_k(\tau)~,
 \ee
where the overdot indicates a conformal time derivative $d/d\tau$.
%Since the interaction between gravitational waves and other
%matters are very weak, in a lot of cases, the right-hand of this
%equation is negligible, so the evolution of RGW is only dependent
%on the scale factor and its time derivative.

The RGW generated during the early inflation stage. Inflation is
an extremely attractive idea to describe the very early universe,
which has received strong support from the observations of CMB
anisotropies and from studies of the large-scale distribution of
galaxy. In this paper, we will only consider the simplest single
field models. This kind of models is enough to account for the
current observations on $n_s$, $\alpha$, and $r$. In the context
of slow-roll inflationary models, the most observables depend on
three slow-roll parameters \cite{slow-roll}
 \be\label{p}
 \epsilon_V\equiv\frac{\Mpl^2}{2}\left(\frac{V'}{V}\right)^2~,
 ~~~~~~
 \eta_V\equiv \Mpl^2\left(\frac{V''}{V}\right)~,
 ~~~~~~
 \xi_V\equiv \Mpl^4\left(\frac{V'V'''}{V^2}\right)~,
 \ee
where $\Mpl\equiv(8\pi G)^{-1/2}=\mpl/\sqrt{8\pi}$ is the reduced
Planck energy. In the following discussion, I will use the unit
$\Mpl\equiv1$ and $\mpl=\sqrt{8\pi}$. $V(\phi)$ is the
inflationary potential, and prime denotes derivatives with respect
to the field $\phi$. Here, $\epsilon_V$ quantifies ``steepness" of
the slope of the potential, $\eta_V$ quantifies ``curvature" of
the potential and $\xi_V$ quantifies the ``jerk". All parameters
must be smaller than one for inflation to occur. The most
important prediction of the inflationary models is the primordial
scalar perturbation power spectrum, which is nearly gaussian and
nearly scale-invariant. This spectrum is always written in the
form
 \be\label{s-p}
 P_S(k)=P_S(k_0)\left(\frac{k}{k_0}\right)^{n_s(k_0)-1+\frac{1}{2}\alpha\ln(k/k_0)}~,
 \ee
where $\alpha\equiv dn_s/d\ln k$, and $k_0$ is some pivot
wavenumber. In this paper, $k_0=0.002$Mpc$^{-1}$ is used. The
observations of WMAP find $P_S(k_0)\simeq2.95\times10^{-9}A(k_0)$
and $A(k_0)=0.813^{+0.042}_{-0.052}$ \cite{map3}. Another key
prediction of inflationary models is that the existence of the
RGW. The primordial power spectrum of RGW is defined by
 \be\label{6}
 P_T(k)\equiv\frac{32Gk^3}{\pi}h^{+}_kh_k~.
 \ee
The strength of the gravitational waves is characterized by the
gravitational waves energy spectrum
 \be\label{26}
 \Omega_{g}(k)=\frac{1}{\rho_c}\frac{d\rho_{g}}{d\ln k}~,
 \ee
where $\rho_c=3H_0^2/8\pi G$ is the critical density and
$H_0=100h~$km s$^{-1}$Mpc$^{-1}$ is the present Hubble constant.
One can relate $\Omega_g$ to the primordial power spectrum by the
formula \cite{boy,before-work}
 \be\label{ome}
 \Omega_{g}(k)=\frac{1}{12H_0^2}k^2P_T(k)T_f^2(k)~,
 \ee
where the transfer function $T_f(k)$ reflects the damping effect
of the gravitational waves when evolving in the expansion
universe. It is convenient to defined a function ${\cal
T}(k)\equiv k^2T_f^2/12H_0^2$, so the strength of RGW becomes
$\Omega_{g}(k)={\cal T}(k)P_T(k)$. In the following sections, we
will discuss ${\cal T}(k)$ and $P_T(k)$, respectively.

\section{The Transfer Function}

In this section, we will discuss three kinds of damping effects:
First we ignore the anisotropy stress in Eq.(\ref{h-evolution}),
and only consider the redshift-suppression effect. So
Eq.(\ref{h-evolution}) becomes
 \be\label{vacuum}
 \ddot{h}_{k}+2\frac{\dot{a}}{a}\dot{h}_k+k^2h_k=0~.
 \ee
This is the evolutive equation of RGW in vacuum, which only
depends on the evolution of the scale factor $a(\tau)$. It is
clear that, the mode function of the gravitational waves behaves
simply in two regimes when evolving in the universe: far outside
the horizon ($k\ll aH$), and far inside the horizon ($k\gg aH$).
When waves are far outside the horizon, the amplitude of $h_k$
keeps constant, and when inside the horizon, they damp with the
expansion of the universe
 \be\label{28}
 h_k\propto \frac{1}{a}  ~~.
 \ee
In the simple cosmic model, the evolution of the universe can be
separated into three stages: the radiation-dominant stage, the
matter-dominant stage, and the vacuum-dominant stage. In this
model, by numerically integrating the Eq.(\ref{vacuum}), one found
the transfer function can be approximately described with a
damping function (for the waves with $k\gg10^{-18}$Hz)
\cite{previous,damp1,damp3}:
 \be
 t_1(k)=\frac{3}{(k\tau_0)^2}\frac{\Omega_m}{\Omega_{\Lambda}}
 \sqrt{1.0+1.36\left(\frac{k}{k_{eq}}\right)+2.50\left(\frac{k}{k_{eq}}\right)^2}~~,
 \ee
where $k_{eq}=0.073\Omega_mh^2$Mpc$^{-1}$ is the wavenumber
corresponding to the Hubble radius at the time that matter and
radiation have equal energy densities. And
$\tau_0=1.41\times10^{4}$Mpc is the present conformal time.
$\Omega_m$ and $\Omega_{\Lambda}$ are the present energy densities
of matter and vacuum, respectively. It is obvious that, when $k\ll
k_{eq}$, which entered the horizon in the matter-dominant or
vacuum-dominant stage, $t_1(k)\propto k^{-2}$, but when $k\gg
k_{eq}$, which entered the horizon in the radiation-dominant
stage, $t_1(k)\propto k^{-1}$, which is for the different
evolution of scale factor in different stages. The factor
$\Omega_m/\Omega_{\Lambda}$ is the effect of accelerating
expansion, which has been discussed in the previous
works\cite{damp3,previous,accele}.

The second is the damping effect of the free-streaming
relativistic particles \cite{damp2}, especially the neutrino,
which can generate the anisotropic stress $\Pi_k$ on the
right-hand of the Eq.(\ref{h-evolution}), when it is the
free-streaming relativistic particles. This effect was first
considered by Weinberg, where the Eq.(\ref{h-evolution}) can be
rewritten as a fairly simple integro-differential equation. The
solution shows that anisotropy stress can reduce the amplitude for
the wavelengths that re-enter the horizon during the
radiation-dominated stage, and the damping factor is only
dependent on the fraction $f$ of the background (critical) energy
density of the free-streaming relativistic particles in the
universe. The effect is less for the wavelengths that enter the
horizon at later time. A lot of works have been done to simplify
this effect, and in Ref.\cite{boy}, the authors found it can be
approximately described by a transfer function $t_2$ for the waves
with $\nu>10^{-16}$Hz (which re-enter the horizon at the
radiation-dominant stage),
 \be\label{t3}
 t_2=\frac{15(14406f^4-55770f^3+3152975f^2-48118000f+324135000)}
 {343(15+4f)(50+4f)(105+4f)(180+4f)}~~.
 \ee
When the wave modes ($10^{-16}$Hz$<\nu<10^{-10}$Hz) re-enter the
horizon, the temperature in the universe is relatively low
($<1$MeV), we are fairly confident that the neutrino is the only
free-streaming relativistic particle. So we choose $f=0.4052$,
corresponding to $3$ standard neutrino species, the damping factor
is $0.80313$. But for the waves with very high frequency
($\nu>10^{-10}$Hz), the temperature of the universe is very high
when they re-enter the horizon, and the value of $f$ is much
uncertain. Thus, the detection of RGW at this frequency offers the
probability of learning about the free-streaming fraction $f$ in
the very early universe.

The third is the effect due to the successive changes in the
relativistic degrees of freedom during the radiation-dominant
stage, here we also call it the effect due to the cosmic phase
transition, which includes the QCD transition, the $e^+e^-$
annihilation, the electroweak phase transition and so on. In an
adiabatic system, the entropy per unit comoving volume must be
conserved \cite{kolb,trans},
 \be
 S(T)=s(T)a^3(T)={\rm constant}, ~~~and~~s(T)=\frac{2\pi^2}{45}g_{*s}(T)T^3,
 \ee
where the entropy density, $s(T)$, is given by the energy density
and pressure; $s=(\rho+p)/T$. Combining it with the expressions of
energy density and pressure in the radiation-dominant universe,
 \be\label{rhop}
 \rho(T)=\frac{\pi^2}{30}g_*(T)T^4, ~~~p(T)=\frac{1}{3}\rho(T)~.
 \ee
one can immediately get the relation
 \be
 \rho\propto g_*g_{*s}^{-4/3}a^{-4},
 \ee
where we have defined the ``effective number of relativistic
degrees of freedom", $g_*$ and $g_{*s}$, following the
Ref.\cite{kolb,trans}. These quantities, $g_*$ and $g_{*S}$, count
the effective number of relativistic species contributing to the
radiation energy density and entropy, respectively. From this
relation, one can find that, if the phase transitions are not
considered, $g_*$ and $g_{*s}$ are all constant, and this relation
becomes to the general case of $\rho\propto a^{-4}$. However, it
does not always hold, as some particles would become
non-relativistic before the others and stop contributing to the
radiation energy density. In other words, the evolution of $\rho$
during the radiation era is sensitive to how many relativistic
species the universe has at a given epoch. As the wave equation of
gravitational waves constraints $(\dot{a}/a)\dot{h}_k$, the
solution of $h_k$ would be affected by $g_*$ and $g_{*S}$ via the
Friedmann equation:
 \be
 \left(\frac{H(\tau)}{H_0}\right)^2=\left(\frac{g_*}{g_{*0}}\right)
 \left(\frac{g_{*s}}{g_{*s0}}\right)^{-4/3}\Omega_r\left(\frac{a}{a_0}\right)^{-4}+
 \Omega_m\left(\frac{a}{a_0}\right)^{-3}+\Omega_{\Lambda},
 \ee
where the subscript $0$ denotes the quantities with the present
values. Here we have considered the Friedmann equation in the
$\Lambda$CDM universe, which is supported by a number of
observations\cite{sn,map1,map3}. Inserting this into
Eq.(\ref{vacuum}), one can numerically calculate the value of
$h_k$ \cite{trans}, which would spend a very long computer time,
since one must integrate that equation from the end of the
inflation to present time, and calculate the waves from
$\nu=10^{-16}$Hz to $0.1$Hz which we are interested. Here we will
give an approximate method, which can describe this effect by a
simple factor $t_3$. We consider the wave $h_k$ with the
wavenumber $k$, which crossed the horizon at $a=a_k$, and the
corresponding Hubble parameter is $H_k$.  So one has
$k=a_kH_k/a_0$. One knows that,when the waves are in the horizon,
$h_k(\tau)\propto 1/a(\tau)$, damping with the expansion of
universe, and when the waves is out the horizon, the $h_k={\rm
constant}$, keeping its initial value. So one can define a factor
 \be\label{f}
 F_k\equiv\frac{{h_k(\tau_0)}}{{h_k(\tau_i)}}=\frac{a_k}{a_0}~~.
 \ee
where $\tau_i$ is the conformal time at the beginning of the
radiation era. During the radiation era, one has
 \be\label{H}
 H=B\left(\frac{g_*}{g_{*0}}\right)^{1/2}\left(\frac{g_{*s}}{g_{*s0}}\right)^{-2/3}
 \left(\frac{a}{a_0}\right)^{-2},
 \ee
where $B=H_0\Omega_r^{1/2}$, is a constant. Using the
Eqs.(\ref{f}), (\ref{H}) and the relation of $k=a_kH_k/a_0$, one
gets
 \be
 F_k=\frac{B}{k}\left(\frac{g_*(T_k)}{g_{*0}}\right)^{1/2}\left(\frac{g_{*s}(T_k)}{g_{*s0}}\right)^{-2/3},
 \ee
where $T_k$ is the temperature when the wave $h_k$ exactly crossed
the horizon. First we can assume $g_{*}=g_{*0}$ and
$g_{*s}=g_{*s0}$ are always satisfied, which is the condition
without changes in the relativistic degrees of freedom during the
radiation era, which follows that $\tilde{F}_k=B/k$. Inserting
this into Eq.(\ref{ome}), one finds the value of $\Omega_g(k)$ is
independent of the wavenumber $k$. However, here we are interested
in the condition with changing $g_{*}$ and $g_{*s}$, and the
factor $t_3$ exactly denotes the difference of these two
conditions, i.e.
 \be
 t_3=\frac{F_k}{\tilde{F}_k}=\left(\frac{g_*(T_k)}{g_{*0}}\right)^{1/2}\left(\frac{g_{*s}(T_k)}{g_{*s0}}\right)^{-2/3},
 \ee
where $g_{*0}=3.3626$ and $g_{*s0}=3.9091$. This factor depends on
the values of $g_*$ and $g_{*S}$ at the early universe. Fig.[1],
presents the evolution of the values of $g_*$ and $g_{*s}$, which
shows that the value of $g_*$ has an obvious accretion when
$T>0.1$MeV. The difference between $g_*$ and $g_{*s}$ only exists
when $T<0.1$MeV. In the expression of $\Omega_g(k)$, this effect
is described by a factor $t_3^{~2}$. Compared with the accurately
numerical calculation, this approximation has the error smaller
that $10\%$. The total transfer function is the combination of
these three effects
 \be\label{t}
 T_f(k)=t_1\times t_2 \times t_3~,
 \ee
where $t_1$ is most important, which approximately shows the
evolution of RGW in the expanding universe. The function of $t_2$
has the most uncertain in this discussion. In the extreme
condition with $f=0$, $t_2=1$ is held, i.e. no damping; and in
another extreme condition with $f=1$, $t_2=0.59$ is held, this
function arrives at its smallest value. In the case of $f=0.4052$,
$t_2=0.80321$ only contributes a damping factor $0.645$ for the
strength of the RGW. The value of $t_3$ is fairly small. For the
extreme condition with $T_k>10^6$MeV ($k>2\times10^{-4}$Hz), one
has $g_*=g_{*s}=106.75$ in the Standard Model ($g_*=g_{*s}=228.75$
in the MSSM), $t_3=0.62$ ($t_3=0.55$ in MSSM) only contributes a
damping factor $0.38$ ($0.30$ in MSSM) to the strength of the RGW.

The experiments which can directly detect the RGW are all
sensitive to the waves with $k\gg k_{eq}$, which re-entered the
horizon during the radiation era. From the previous discussion,
one can get a simple expression of all these damping effects:
 \[
 {\cal T}(k)=\left(\frac{15}{8k_{eq}^2H_0^2\tau_0^4}\right)
 \left(\frac{\Omega_{m}}{\Omega_{\Lambda}}\right)^2
 \left(\frac{g_*(T_k)}{g_{*0}}\right)\left(\frac{g_{*s}(T_k)}{g_{*s0}}\right)^{-4/3}\\\\
 \]
 \be
 \times\left(\frac{15(14406f_k^4-55770f_k^3+3152975f_k^2-48118000f_k+324135000)}
 {343(15+4f_k)(50+4f_k)(105+4f_k)(180+4f_k)}\right)^2~.
 \ee
where $f_k$ is the value of the function $f$ when wave $h_k$
exactly crossed the horizon. This function is dependent on the
values of $\Omega_{\Lambda}$, $\Omega_{m}$, the present energy
densities of vacuum and matter, respectively; the value of
$\tau_0$, the age of the universe; the value of $H_0$, the present
Hubble constant; the values of $g_{*}(T_k)$ and $g_{*s}(T_k)$, the
effective number of relativistic degrees of freedom; and the
fraction $f_k$ of the background (critical) energy density of the
free-streaming relativistic particles in the universe. So this
function includes abundant cosmic information. Using this, the
strength of RGW becomes
 \be
 \Omega_g(k)=P_T(k){\cal T}(k)~.
 \ee
Here we are interested in the waves with $\nu=0.1$Hz, which is the
sensitive frequency of laser interferometers, BBO and DECIGO.
Choosing the cosmic parameters $h=0.72$, $\Omega_m=0.27$,
$\Omega_{\Lambda}=0.73$, $g_*=g_{*s}=106.75$ and $f_k=0$, one gets
 \be\label{o2}
 {\cal T}(k)=4.15\times10^{-7},~~~~\Omega_g(k)=4.15\times10^{-7}P_T(k)~.
 \ee

\section{The Primordial Power Spectrum of RGW}

The primordial spectrum of RGW is always described in a simple
form
 \be\label{7}
 P_T(k)=P_T(k_0)\left(\frac{k}{k_0}\right)^{n_t(k_0)+\frac{1}{2}\alpha_t\ln(k/k_0)}~,
 \ee
where $n_t(k)$ is the tensor spectral index, and $\alpha_t\equiv
dn_t/d\ln k$ is its running. In the single-field inflationary
models, a standard slow-roll analysis gives the below relations of
observable quantities  and slow-roll parameters,
 \be\label{relation}
 n_t=-\frac{r}{8}~,
 ~~~~~~
 \alpha_t=\frac{r}{8}\left[\left(n_s-1\right)+\frac{r}{8}\right]~,
 ~~~~~~
 r=\frac{8}{3}(1-n_s)+\frac{16}{3}\eta_V~,
 \ee
where $r(k)\equiv P_T(k)/P_S(k)$, is the so-called tensor-scalar
ratio. These formulae relate $n_t$ and $\alpha_t$ to the other two
functions $n_s$ and $r$, which are easily to observe. But the
relation between $r$ and $n_s$ is dependent on $\eta_V$, which
depends on the specific inflationary potential. Inserting these
into Eq.(\ref{7}), one gets
 \be\label{pt1}
 P_T(k)=P_S(k_0)\times r\times\left(\frac{k}{k_0}\right)^
 {-\frac{r}{8}+\frac{r}{16}\left[\left(n_s-1\right)+\frac{r}{8}\right]\ln(k/k_0)}~,
 \ee
where $r$ denotes the tensor-scalar ratio at $k=k_0$, i.e.
$r\equiv r(k_0)$ , which is also held in the following sections.
So the primordial spectrum of RGW only depends on $n_s$ and $r$.
The recent constraints come from the observations of three-year
WMAP \cite{map3}, which are
 \be\label{sdss}
 n_s=0.951^{+0.015}_{-0.019}~(68\% ~C.L.)~,~~~~r<0.28 ~(95\% ~C.L.)~.
 \ee
Using the Eq.(\ref{o2}), one gets
 \be\label{appro}
 \Omega_g(k)=9.98\times10^{-16} r \left(\frac{k}{k_0}\right)^
 {-\frac{r}{8}+\frac{r}{16}\left[\left(n_s-1\right)+\frac{r}{8}\right]\ln(k/k_0)}~,
 \ee
where we have chosen $A(k_0)=0.813$. We have plotted the function
$\Omega_g$ (denoting $\Omega_g\equiv\Omega_g(k_1)$, and
$k_1=0.1$Hz) dependent on $r$ in Fig.[2], where $n_s=0.951$ is
used. This result is consistent with our previous work, a larger
$r$ leads to a larger $\Omega_g$. We well know that the formula in
Eq.(\ref{7}) is a very good approximation when the wavenumber $k$
is not much larger (or smaller) than $k_0$. But it may be not a
good approximation at $k_1$, which is more than $16$ order that
$k_0$. So it is necessary to numerically calculate $P_T(k)$, but
it is not easy for the exactly numerically calculation, since one
must calculate the spectrum in a very wide range in wavenumber
(larger than $16$ order), and for each $k$, one must integrate it
from the initial condition to the end of the inflation. In this
section, we will use a semi-numerical method to calculate the
primordial power spectrum of RGW. We will introduce this method by
discussing two kinds of inflationary models, which all satisfy the
current constraints of $n_s$, $\alpha$ and $r$.

First we consider the model with potential (\textbf{Mod.1.1})
$V(\phi)=\Lambda^4(\phi/\mu)^2$, which belongs to the large-field
model, and predicts a fairly larger $r$ \cite{cata}. From
Eq.(\ref{p}), one gets
 \be
 \epsilon_V=\frac{2}{\phi^2},~~~~\eta_V=\frac{2}{\phi^2},~~~~\xi_V=0,
 \ee
where we have denoted $\Mpl\equiv1$. So the slow-roll condition
requires that $\phi\gg\sqrt{2}$, the so-called large-field model.
At the end of inflation, $\epsilon_V=1$ is satisfied, which leads
to $\phi_{end}=\sqrt{2}$. In the initial condition, one has
\cite{slow-roll}
 \be\label{para}
 n_s-1=-6\epsilon_V+2\eta_V,~~~~r=16\epsilon_V,~~~~P_S(k_0)=\frac{V}{24\pi^2\epsilon_V},
 \ee
which follows that
 \be
 \phi_{ini}=\sqrt{8/(1-n_s)},
 ~~~~~r=4(1-n_s)~~~~\Lambda^4/\mu^2=0.75\pi^2P_S(k_0)(1-n_s)^2.
 \ee
Inserting these into the Hamilton-Jacobi formula,
 \be\label{hj}
 2\left[H'(\phi)\right]^2-3H^2(\phi)=-V(\phi)~,
 \ee
one can immediately get the function $H(\phi)$ by the numerical
calculation. We define the e-folds number $N$, and the scale
factor is $a=a_{ini}e^N$. When $k_0$ crossed the horizon, we set
the scale factor $a=a_{ini}=1$ i.e. $N=0$. The relation of $N$ and
$\phi$ is
 \be\label{phiN}
 \frac{d\phi}{dN}=-2\frac{H'}{H}~,
 \ee
where $H$ is the Hubble parameter during inflation, and $H'\equiv
dH/d\phi$. One can define a Hubble slow-roll parameter
$\epsilon\equiv2(H'/H)^2$, so the primordial power spectrum of RGW
is (to the first slow-roll order) \cite{lyth}
 \be\label{pt2}
 P_T(k)=\left.\frac{2}{\pi^2}\left[1-\frac{c+1}{4}\epsilon\right]H^2\right|_{k=aH},
 \ee
where $c=4(\ln 2+\gamma)-5\simeq0.0814514$ (with $\gamma$ the
Euler-Mascheroni constant) is a constant. Using the Eqs.(\ref{hj})
and (\ref{phiN}), one can numerically calculate $H(N)$. Inserting
it into Eq.(\ref{pt2}), one can get the primordial spectrum of
RGW, at the same time the total e-folds $N$ is also got. The value
of $\Omega_g$ is also got by using the Eq.(\ref{o2}). We have
plotted the values of $r$ and $\Omega_g$ in Fig.[2], where we have
chosen $n_s\in[0.94,~0.98]$. It is easily found the value of $r$
is in the range $r\in[0.08,~0.24]$. Compared with the approximate
formula (\ref{appro}), the numerical value is much smaller: When
$n_s=0.951$, the value is only one third of the approximate value.

Then let us consider another model
$V(\phi)=\Lambda^4[1-(\phi/\mu)^2]$, which belongs to the
small-field model, and predicts a very small $r$ \cite{cata}. From
Eq.(\ref{p}), one gets
 \be
 \epsilon_V=\frac{1}{2}\left[\frac{2x/\mu}{1-x^2}\right]^2,~~~~\eta_V=\frac{2/\mu^2}{x^2-1},~~~~\xi_V=0,
 \ee
where $x\equiv\phi/\mu$. At the end of inflation, $\phi_{end}=\mu$
i.e. $x_{end}=1$, where $V=0$ is satisfied. The initial value of
$x$ must be very small to account for the slow-roll condition.
Since it can not be got from the observed $n_s$ and $P_S(k_0)$, we
must set it before the calculation. First let us consider the
model with $x_{ini}=0.1$ (\textbf{Mod.2.1}), using the
Eq.(\ref{para}), one immediately gets
 \be
 \mu^2=4.89746/(1-n_s),~~~~r=0.06667(1-n_s),~~~~\Lambda^4=0.99693(1-n_s)P_S(k_0);
 \ee
Second we consider the model with $x_{ini}=0.2$
(\textbf{Mod.2.2}), which follows that
 \be
 \mu^2=6.07693/(1-n_s),~~~~r=0.22857(1-n_s),~~~~\Lambda^4=3.52486(1-n_s)P_S(k_0);
 \ee
The third model has $x_{ini}=0.3$ (\textbf{Mod.2.3}), which
follows that
 \be
 \mu^2=7.72854/(1-n_s),~~~~r=0.45(1-n_s),~~~~\Lambda^4=7.32086(1-n_s)P_S(k_0).
 \ee
Then using the Hamilton-Jacobi formula in Eq.(\ref{hj}) and the
relation of $N$ and $\phi$, here it becoming $d\phi/dN=2H'/H$, one
can also get the function $H(N)$. Using the formula (\ref{pt2}),
the values of $r$, $P_T(k)$, $\Omega_g(k)$ and $N$ are also got,
which have been plot in Figs.[2] and [3]. From Fig.[3], one finds
a larger $n_s$ leads to a larger $N$, which is held for all these
four inflationary models. When $n_s=0.951$, $N=41.96$ for the
Mod.1.1, and $N=62.47$ for the Mod.2.3, which are in the region of
$N\in[40,~70]$, and acceptable \cite{N}. But for the Mod.2.1,
$N=97.90$, and for the Mod.2.2, $N=74.59$, which are too large to
suitable. From Fig.[2], one finds that when $n_s\in[0.94,~0.98]$,
$r<0.02$ is satisfied for the Mod.2.1, Mod.2.2, Mod.2.3, the very
small values. And values of $\Omega_g$ are exactly same with the
approximate results. So one gets a conclusion: when $r$ is small,
the formula (\ref{appro}) is a very good approximation, but when
$r$ is larger ($r>0.1$), the approximate formula (\ref{appro}) is
not very good, and the numerical calculation is necessary.

\section{Conclusion and Discussion}

Inflation has received strong supports from the observations of
the CMB and LSS. As a key prediction of inflationary models, the
detection of RGW would provide incontrovertible evidence that
inflation actually occurred and set strong constraints on the
dynamics of inflation. A lot of experiments are under development
for the RGW detection, which mainly include two kinds: The CMB
experiments, including Planck, Clover, and others, and the laser
interferometers, including BBO, DECIGO and so on. For
investigating the detection abilities of these two kinds of
experiments, it is convenient to study the distribution of the
inflationary models in the $r-\Omega_g$ diagram. So it is
necessary to accurately calculate the RGW at all frequency range.
In this paper, we improved the previous calculation in two
aspects: First, we studied the transfer function after considering
the redshift-suppression effect, the accelerating expansion
effect, the damping effect of free-streaming relativistic
particles, and the damping effect of cosmic phase transition, and
gave a simple approximate formula of the transfer function, which
applies to the waves with $k>k_{eq}$. This function depends on the
values of the cosmic parameters: $\Omega_{m}$, $\Omega_{\Lambda}$,
$H_0$, $k_{eq}$, $\tau_0$, $g_{*}$, $g_{*s}$, and $f_k$. Second,
we have developed a numerical method to calculate the primordial
power spectrum of RGW, especially at high frequency, where the
observed constraints on $n_s$ and $P_S(k_0)$ and the
Hamilton-Jacobi equation are used. We applied this method to two
kinds of inflationary models, which all satisfy the current
constraints on $n_s$, $\alpha$ and $r$.

From Fig.[3], one can find: in all these inflationary models, a
larger $n_s$ follows a larger $N$. For the first kind of model,
when $n_s>0.97$, the value of $N>70$ is satisfied, which is
unsuitable. To account for the constraint of $N\in[40,~70]$, $n_s$
can only be in a very narrow region $n_s\in[0.948,~0.970]$. For
the second kind of model, the initial conditions of $x_{ini}=0.1$
and $x_{ini}=0.2$ are not acceptable, which predict too large
e-folds. The condition of $x_{ini}=0.3$ is suitable, which
predicts $N=62.47$ when $n_s=0.951$. But to account for the
constraint $N<70$, $n_s<0.956$ must be satisfied. From Fig.[2],
one found that, for the Mod.1.1, when $n_s\in[0.94,~0.98]$, the
value of $r$ is in the region $r\in[0.08,~0.24]$, which are mostly
in the sensitive region of Planck satellite. The value of
$\Omega_g$ is in the region of
$[5.6\times10^{-17},~2.2\times10^{-18}]$. In the most region of
$n_s$, a larger $n_s$ follows a smaller $r$, and corresponds to a
larger $\Omega_g$, which is an unexpected result. This is an
obvious difference from the result of the approximate formula.
When $n_s=0.951$, $\Omega_g=1.3\times 10^{-17}$, which is in the
sensitive region of ultimate DECIGO, but beyond the sensitive
limit of BBO. This value is only one third of the value from the
approximate formula in Eq.(\ref{appro}). For the Mod.2.1, Mod.2.2
and Mod.2.3, a larger $n_s$ follows a smaller $r$ and a smaller
$\Omega_g$. These models predicted a very small $r$, when
$n_s\in[0.94,~0.98]$, $r<0.02$ is always satisfied, and the value
of $\Omega_g$ is exactly same with the the value from the
approximate formula in Eq.(\ref{appro}). For the Mod.2.3, which
predicted an acceptable e-folds, the values of $r$ are all in the
sensitive region of Clover, but beyond which of Planck; the values
of $\Omega_g$ are all in the sensitive region of ultimate DECIGO,
but beyond which of BBO.

\begin{figure}
\centerline{\includegraphics[width=12cm]{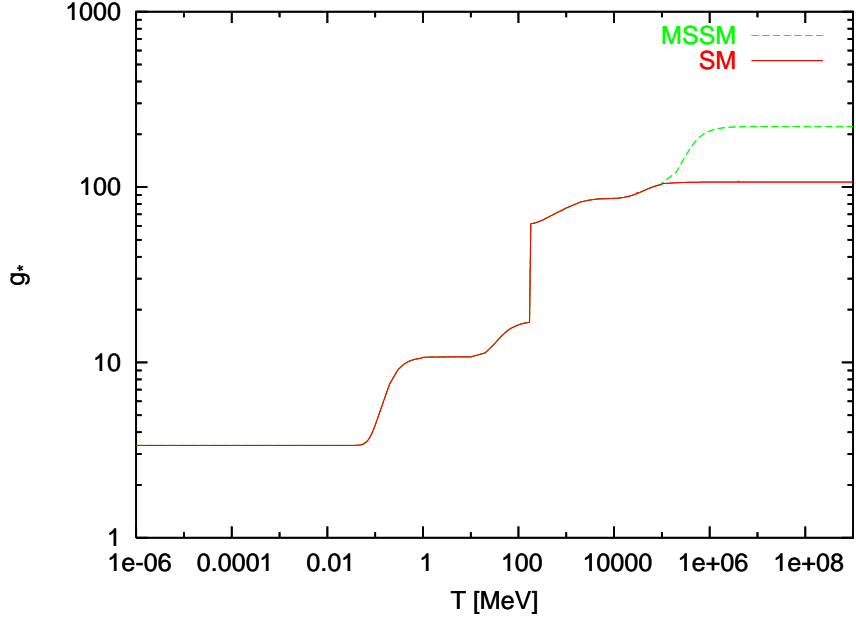}}
 \caption{\small Evolution of $g_{*}$ with the temperature. The solid and dash lines
 represent $g_{*}$ in the Standard Model (SM) and in the Minimal extension of Supersymmetric Standard Model (MSSM),
 respectively. At the energy scales below $\sim0.1$MeV, $g_{*}=3.3626$ and $g_{*s}=3.9091$;
 $g_{*}=g_{*s}$ otherwise. This figure comes from the Ref.\cite{trans}}
\end{figure}

\begin{figure}
\centerline{\includegraphics[width=15cm]{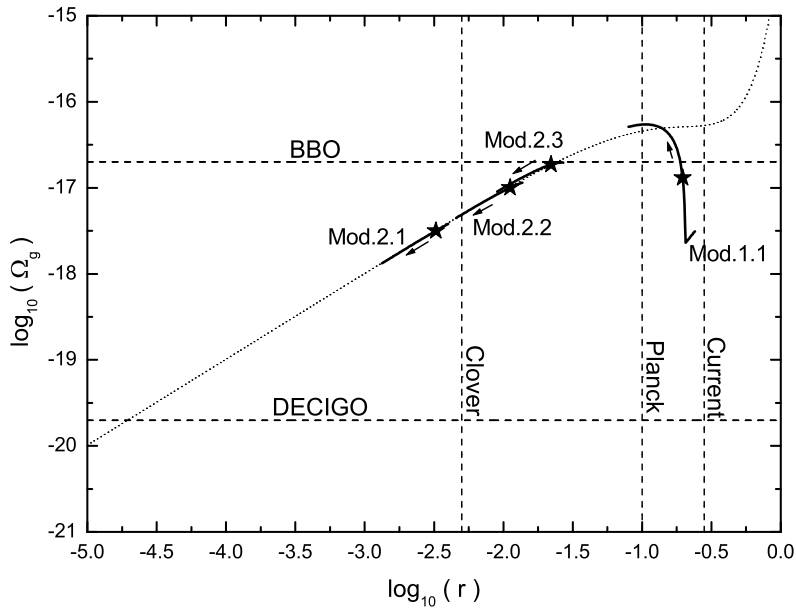}}
 \caption{\small The inflationary models in the $r-\Omega_g$ diagram. $r$ is the tensor-scalar ratio
 at $k=0.002$Mpc$^{-1}$, and $\Omega_g$ is the strength of RGW at $k=2\pi\times0.1$Hz. The dot line
 represents the curve of the approximate formula in Eq.(\ref{appro}) and $n_s=0.951$.
 The vertical (dot) lines from right to
 left are the sensitive limit curves of current observations, Planck and Clover,
 respectively. The horizontal (dot)
 lines from up to down are the sensitive limit curves of BBO and ultimate DECIGO, respectively.
 The solid lines are the predicted curves of the
 inflationary models, where $n_s\in[0.94,~0.98]$, and the arrows
 denote the direction with increasing $n_s$. The stars denote
 the models with $n_s=0.951$.
  }
\end{figure}

\begin{figure}
\centerline{\includegraphics[width=15cm]{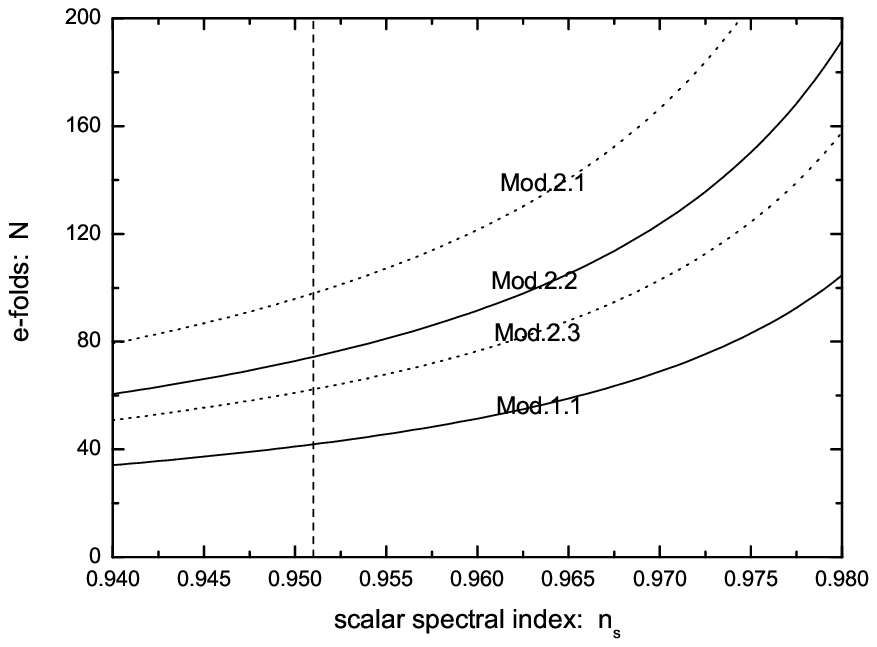}}
 \caption{\small The evolution of the value of the e-folds $N$ with the scalar spectral index $n_s$ for the inflationary models.
 The black dot line denotes the curve with $n_s=0.951$. }
\end{figure}

~

\textbf{ACKNOWLEDGMENT}: The author thanks Yuki Watanabe for
useful help.

%%%%%%%%%%%%%%%%%%%%%%%%%%%%%%%%%%%%%%%%%%%%%%%%%%%%%%%%%%%%%%%%%%%%%%%%%%%%%%%%%%%
%%%%%%%%%%%%%%%%%%%%%%%%%%%%%%%%%%  BIBLIOGRAPHY  %%%%%%%%%%%%%%%%%%%%%%%%%%%%%%%%%%%%%%%%
%%%%%%%%%%%%%%%%%%%%%%%%%%%%%%%%%%%%%%%%%%%%%%%%%%%%%%%%%%%%%%%%%%%%%%%%%%%%%%%%%%%

\end{document}